\documentclass[fleqn,usenatbib]{mnras}

\usepackage{newtxtext,newtxmath}

\usepackage[T1]{fontenc}

\DeclareRobustCommand{\VAN}[3]{#2}
\let\VANthebibliography\thebibliography
\def\thebibliography{\DeclareRobustCommand{\VAN}[3]{##3}\VANthebibliography}

\usepackage[english]{babel}
\usepackage{graphicx}	
\usepackage{amsmath}	
\usepackage{booktabs}   

\usepackage{colordvi}
\newcommand{\ms}{MS}
\newcommand{\rms}{\textit{Reduced Main Sample}}
\newcommand{\n}{\texttt{Null}}
\newcommand{\nd}{\texttt{ND}}
\newcommand{\tm}{\texttt{TM}}

\title[PRF selection of QSO candidates]{The Probabilistic Random Forest applied to the selection of quasar candidates in the QUBRICS Survey}

\author[Guarneri  et al.]{
\parbox[t]{\textwidth}{
Francesco Guarneri$^{1,2}$\thanks{E-mail: francesco.guarneri@inaf.it},
Giorgio Calderone$^{2}$,
Stefano Cristiani$^{2,3,4}$,
Fabio Fontanot$^{2}$,
Konstantina Boutsia$^{5}$,
Guido Cupani$^{2}$, 
Andrea Grazian$^{6}$,
Valentina D'Odorico$^{2,3,7}$
}
\vspace*{6pt}\\
$^{1}$ Dipartimento di Fisica, Sezione di Astronomia, Università di Trieste, via G.B. Tiepolo 11, I-34131, Trieste, Italy \\
$^{2}$ INAF--Osservatorio Astronomico di Trieste, Via G.B. Tiepolo, 11, I-34143 Trieste, Italy \\
$^3$ IFPU--Institute for Fundamental Physics of the Universe, via Beirut 2, I-34151 Trieste, Italy \\
$^4$ INFN-National Institute for Nuclear Physics,  
via Valerio 2, I-34127 Trieste, Italy \\
$^5$ Las Campanas Observatory, Carnegie Observatories, 
Colina El Pino, Casilla 601, La Serena, Chile\\
$^6$ INAF--Osservatorio Astronomico di Padova,
Vicolo dell'Osservatorio 5, I-35122 Padova, Italy \\
$^7$ Scuola Normale Superiore, 
P.zza dei Cavalieri, I-56126 Pisa, Italy\\
}

\date{Accepted XXX. Received YYY; in original form ZZZ}

\pubyear{2021}

\begin{document}
\label{firstpage}
\pagerange{\pageref{firstpage}--\pageref{lastpage}}
\maketitle

\begin{abstract}
The number of known, bright ($i<18$), high-redshift ($z>2.5$) QSOs in the Southern Hemisphere is considerably lower than the corresponding number in the Northern Hemisphere due to the lack of multi-wavelength surveys at $\delta<0$. Recent works, such as the QUBRICS survey, successfully identified new, high-redshift QSOs in the South by means of a machine learning approach applied on a large photometric dataset.
Building on the success of QUBRICS, we present a new QSO selection method based on the Probabilistic Random Forest (PRF), an improvement of the classic Random Forest algorithm. The PRF takes into account measurement errors, treating input data as probability distribution functions: this allows us to obtain better accuracy and a robust predictive model. We applied the PRF to the same photometric dataset used in QUBRICS, based on the SkyMapper DR1, Gaia DR2, 2MASS, WISE and GALEX databases. The resulting candidate list includes $626$ sources with $i<18$. We estimate for our proposed algorithm a completeness of $\sim$84\% and a purity of $\sim78\%$ on the test datasets. Preliminary spectroscopic campaigns allowed us to observe 41 candidates, of which 29 turned out to be $z>2.5$ QSOs.

The performances of the PRF, currently comparable to those of the CCA, are expected to improve as the number of high-z QSOs available for the training sample grows: results are however already promising, despite this being one of the first applications of this method to an astrophysical context.
\end{abstract}




\section{Introduction}
Luminous quasars, especially at high redshift, 
play the paramount role of cosmic beacons for a variety of
studies on the formation and evolution of galaxies and supermassive black holes (SMBH), Dark Matter, primordial elements, reionization, cosmological parameters, fundamental constants and General Relativity. However, finding quasars at high-z is not a trivial task, due to their relative scarcity with respect to other sources with the same apparent luminosity. The advent of the SDSS survey \citep[e.g.][]{ref:DR16} has represented a significant improvement in this respect, at least in the Northern Hemisphere. At present, the SDSS has delivered more than $10^5$ \citep[][]{ref:Lyke20sdss16q} spectroscopically confirmed QSOs at $0<z<6.5$, with a large fraction at absolute magnitudes $M_{1450}\le -26$. 

In the Southern Hemisphere, due to the lack of wide multi-wavelength surveys at $\delta\le 0^\circ$, the situation used to be significantly less favorable. Comparing QSO surface densities \citep[e.g.][]{ref:Veron10} in different parts of the sky, of the 22 known QSOs with $z>3$ and $V<17$, only 5 have been found at $\delta < 0^\circ$, and all the 3 QSOs with $V<16$ are in the North. Besides, recent studies point out that even an exquisite survey as the SDSS can suffer from incompleteness due to color selection \citep[e.g.][]{ref:Fontanot07, ref:Sch19a}. As a
consequence, also in the Northern Hemisphere high-z QSO densities could be biased towards lower numbers due to the adoption of efficient but relatively incomplete selections.

In \citet[]{ref:Calderone19} and
\citet[]{ref:Boutsia20} we presented the first results of the QUBRICS survey, aimed at finding $z \geq 2.5$ QSOs at bright $i$-band magnitudes ($i\le 18$) in the Southern Hemisphere, taking advantage of the recent availability of new multi-wavelength public databases. The candidate selection in QUBRICS has been based on the Canonical Correlation Analysis \citep[CCA,][]{ref:CCA} and its success rate in finding $z > 2.5$ QSOs is estimated to be around 70\%, with the predominant contaminants being lower-z QSO at $z < 2.5$. Its completeness, evaluated against the presently known bright QSOs at $z > 2.5$, turns out to be of the order of 90\% \citep{ref:Calderone19}.

In the present paper we explore the possibility to use other selection methods in order to further increase the purity and completeness of the QUBRICS sample, and to fully exploit the information content of the multi-band photometric databases on which QUBRICS is based.
In particular, in this paper we will present and discuss a selection procedure based on the Probabilistic Random Forest \citep[PRF, ][]{ref:reisPRF}, an improvement of the Random Forest that makes it possible to properly include measurement errors in the predictive model and to handle missing data in the dataset.

The paper is organized as follows: in section \ref{sec:ml} we will briefly describe the Random and Probabilistic Random Forest; section \ref{sec:samplePrep} will describe how the initial dataset has been prepared, while in section \ref{sec:results} the results obtained from our tests will be presented; in section \ref{sec:characterization} we will attempt to characterize the spectroscopic sample and compare our results with those obtained with the CCA method, while section \ref{sec:spectroscopy} will describe the results of a small spectroscopic campaign. Conclusions are drawn in section \ref{sec:conclusions}.

\section{Machine learning techniques} \label{sec:ml}
Modern astronomical datasets are rapidly growing both in size and complexity, thanks to recent multi-wavelength and multi-epoch surveys such as the SDSS \citep[][]{ref:DR16}, GAIA \citep{ref:GaiaEDR3}, DES \citep[][]{ref:DESDr2} or Pan-STARRS \citep{ref:panstarrs}; machine learning (ML) methods are becoming increasingly popular as automatic tools to perform a variety of tasks on these databases \citep{ref:Baron19MLReview}.

Machine learning (ML) techniques are generally classified into two broad groups/categories: supervised and unsupervised methods. The former are used to map a set of features to a target \textit{label} or quantity, which is provided by a third party actor (another algorithm or a human expert) while the latter are used to infer existing relationships in the dataset, without relying on external labels.

Given a dataset, individual elements are called \textit{objects}, and data associated with a single object \textit{features}. As a practical example, a dataset may be a large photometric collection, where each object is an observed source and each feature is a magnitude measurement. Target labels and quantities differ depending on the specific task, as supervised learning can be used both for classification and regression: in the first scenario (classification) the label is discrete; in the second (regression) it is continuous. Examples for the two cases are, respectively, the classification of a source as a star or a quasar and the estimate of the redshift given a number of photometric measurements. Supervised methods also have model parameters and hyper-parameters: the former are learnt from the data which the model is trained on and are required in the prediction stage; the latter are instead set by the user and fine-tuned to obtain the best performances out of a ML algorithm.

To assess the capabilities of a ML algorithm, it is common practice to subdivide the available dataset into three sub-samples: a training, validation and testing dataset. The first sample is used to train the algorithm; the validation dataset is used to find the optimal hyper-parameters for the specific task of interest and to gain finer control of the learning process (e.g. to prevent overfitting). The last dataset is finally used to estimate the predictive capabilities of the algorithm, as the learnt model is applied to an unseen dataset. Training, validation and testing sets should be independent to obtain an unbiased evaluation of the performances of the algorithm. An alternative approach, especially useful in case of a limited dataset, is the $k$-fold cross validation: the original dataset is split in two parts, a training/validation and testing dataset. Training and validation are carried out at the same time: the training dataset is split in $k$ subsets; in turn, one of these $k$ subsets is used as validation set, while the algorithm is trained on the remaining $k-1$ subsets. This allows to perform the validation process without requiring additional subdivisions in the base dataset.

Despite their widespread use and proved success in Astronomy \citep[e.g][and references therein]{ref:carrasco15}, machine learning algorithms in general are not designed to deal with datasets in which the features have different uncertainties.
However, the performances of ML algorithms strongly depend on the signal to noise of the input data \citep{ref:reisPRF}, suggesting that noise and measurement errors play an important part in the learning and predictive process. Available algorithms can be modified to account for uncertainties during the training process, but simple methods are unsuited to extract all available information: for instance, in a Random Forest algorithm (which will be described in section \ref{sec:RF}) uncertainties in the dataset can be used as additional features; the association between measurements and errors is however indirect, as there is not an explicit probability distribution function involved.

An alternative approach, the Probabilistic Random Forest (PRF), has been recently developed by \citet{ref:reisPRF}, who modified the Random Forest technique to directly account for measurement errors.

\subsection{The original Random Forest} \label{sec:RF}
The Random Forest is an ensemble learning method - an algorithm which uses multiple learning algorithms to obtain better predictive performance than any of the constituent learning algorithm alone - that operates by creating a large number of decision trees during the training process \citep[][]{rif:Breiman2001}.

Decision trees are predictive models described by a tree-like graph, used both for classification and regression tasks (\citet{ref:reisPRF}. Examples of both employments can be found in \citet{ref:RFClassifier, ref:RFRegression}; in the following, however, we will focus on classification tasks.

Each decision tree is built out of a set of consecutive nodes, and each node is a condition on a feature of the dataset. Conditions are in the form of decision branch:
\begin{equation}
	X_i > X_{i, th},
	\label{eq:RFCondition}
\end{equation}
in which $X_i$ is the $i^{th}$ feature for objects in the dataset and $X_{i, th}$ is a threshold value. Both the feature and the threshold value for a node are determined during the training process based on the minimization of a cost function, commonly based on the Gini impurity. The Gini impurity of a given subset is the probability of misclassifying an object, if it is assigned a label randomly drawn from the label distribution of that same subset \citep[][]{ref:BreimannGiniImpurity}. 

We consider as an example a simple two-class (A, B) classification task: the training process starts with the whole training set and a single node, the root of the tree. The algorithm searches for the feature and threshold value that produces the best split, i.e. the one that minimized the aforementioned cost function, determining the condition for the root node.
Objects in the training set are then split in two subset, one for which eq. \ref{eq:RFCondition} is satisfied, one for which eq. \ref{eq:RFCondition} is not. For both of these, a new best-splitting feature is searched: the process continues iteratively as long as the combined impurity of the resulting two child nodes is lower than the impurity of the parent node. If this condition is not satisfied the current node becomes a terminal node (\textit{leaf}) which does not carry a condition but rather a label: this is determined according to the most common label in the subset associated to the terminal node itself. During the classification process an unlabeled object is propagated along a decision tree according to its feature values and is finally classified based on the terminal node it reaches. An example of a decision tree can be found, for instance, on the scikit learn \citep[][]{ref:scikit} website\footnote{\href{https://scikit-learn.org/stable/modules/tree.html}{https://scikit-learn.org/stable/modules/tree.html}}.

A simple, unpruned, decision tree is not limited in its size, and shows perfect performances on the training set, while typically showing worse performance when applied on new, unseen data: this behaviour is generally referred to as overfitting \citep[][]{rif:Breiman2001}. The random forest mitigates the issue using numerous decision trees and introducing randomness in the training process. This is usually done using two complementary approaches: each tree is trained on a randomly extracted subset of the original dataset (a technique which is also called bootstrap), and for each node the best splitting feature is chosen from a random subset of all available features; the dimension of the subset is one of the hyper-parameters set by the user.
During the prediction process each tree independently classifies each object; the final class is determined by majority vote, i.e. the most common class among all trees is chosen.

\subsection{Probabilistic Random Forest} \label{chap:prf}
The Probabilistic Random Forest \citep[PRF, ][]{ref:reisPRF} is an improvement of the original RF designed to properly handle measurement errors. The main difference between the RF and the PRF consists in the treatment of input data: a 'classic' RF algorithm maps feature and labels, while, on the other hand, the PRF also takes feature and labels uncertainties ($\Delta$X and $\Delta$y) into account in order to identify the optimal mapping function.

Uncertainties arise both from measurements ($\Delta X$) and classification labels ($\Delta y$). In the PRF implementation, the two are treated quite differently: features are considered as probability distribution functions (PDF), with expectation value equal to the feature value and variance equal to the associated error squared. On the other hand, labels are treated as probability mass functions (i.e. discrete density functions): each object has a fixed chance of belonging to each class.

\begin{figure}
	\centering
	\includegraphics[width=0.95\linewidth]{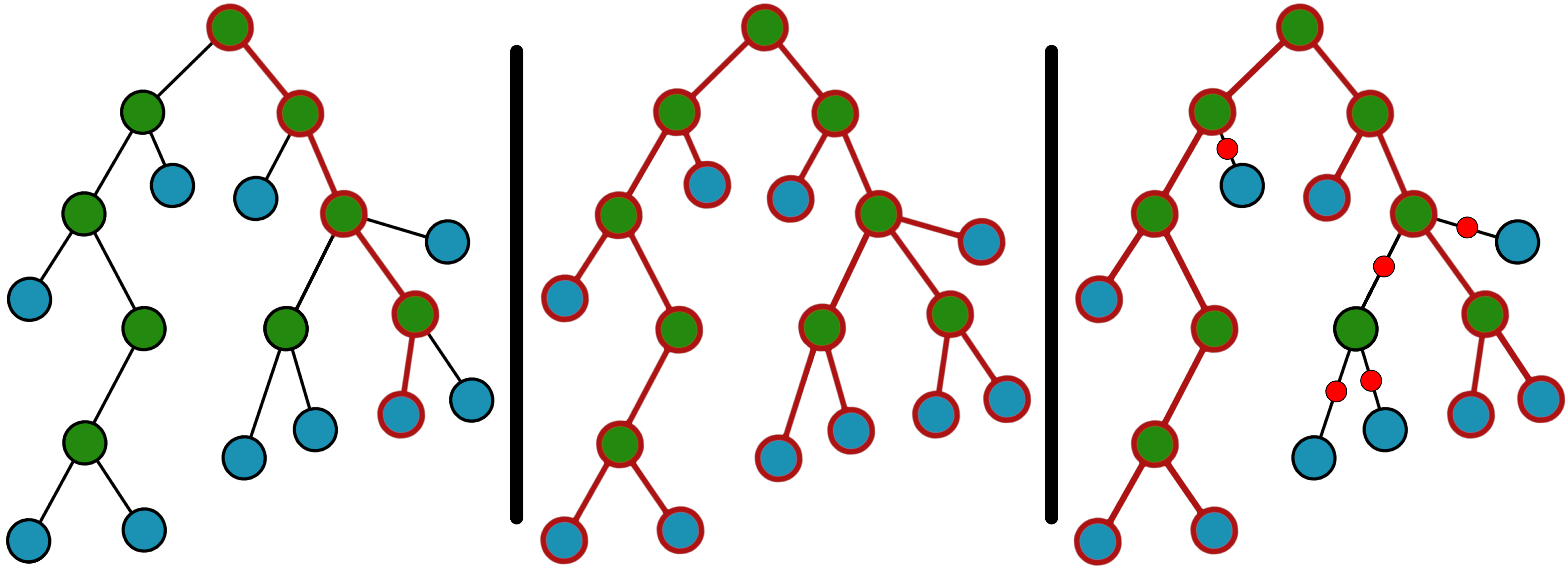}
	\caption{An object propagates through a decision tree. Terminal nodes (leaves) are light blue coloured. The path of an object along a tree is marked by red lines, black lines show all possible paths. The left panel represents the propagation in the classic RF approach: an object propagates either to the left or right node for each split. The middle panel shows an ideal PRF model: an object propagates along the whole tree, reaching all terminal nodes at the same time, and each object may reach several leaves (although with different probabilities). The right panel shows a "pruned" PRF (i.e. with a set probability threshold): red dots represent nodes which can't be reached due to low probabilities associated with those splits. Adapted from \citet{ref:reisPRF}.}
	\label{fig:prf}
\end{figure}

This simple change has an important effect on decision trees: in a RF an unlabelled object in a given node propagates either to the subsequent right or left node. In the PRF, instead, each object propagates into both nodes with a given probability (fig. \ref{fig:prf}); the probability of propagating to the left or right branch are given by the cumulative distribution function for a particular feature; in the current PRF implementation\footnote{\href{https://github.com/ireis/PRF}{https://github.com/ireis/PRF}} the PDF is chosen for all objects as a Gaussian, but the choice can be arbitrary. Moreover, as all objects propagate along the whole tree, all leaves contribute to the classification solution of each object.

In principle, any object can always propagate to the next node, even if the probability to do so is small. To optimise the algorithm a probability threshold is introduced: this is implemented in the PRF  as an adjustable parameter (\texttt{keep\_proba}), with a default value of 0.05 (i.e. an object does not propagate to subsequent nodes if the probability to do so is less than 5\%). 

\noindent The PRF has several advantages over the classic RF:
\begin{itemize}
	\item noise robustness: \citet{ref:reisPRF} tested various noise injections in both the training and testing data, finding that in almost all cases the PRF outperforms the original RF.
	Improvements in the performance of the algorithm depend on the noise characteristics: noise which produces a clear distinction in objects with poorly and well measured features leads to negligible improvements; complex noise, that does not result in a clear distinction between feature quality leads to a greater boost in the classification abilities. This is even more noticeable when the noise is different in the training and testing dataset, which is a possible occurrence in astronomy (for instance when measurements are taken from different catalogues);
	
	\item missing values: these are rather common in astronomical datasets, and sometimes many objects are missing measurements for at least one of the selected features. The PRF can naturally handles missing data: an object with a non-measured feature will just propagate both to the left and the right of a node with 50\% probability.
\end{itemize}

\section{The Preparation of the Sample: the QUBRICS survey} \label{sec:samplePrep}
The PRF needs a training set large enough to produce a robust predictive model. The dataset should include sources of interest - high redshift QSOs - together with those that should be excluded by the selection process: in our case typically non-active galaxies, stars and low-redshift QSOs. 

In this work, we have used the same dataset described in the papers by \citet{ref:Calderone19} and \citet{ref:Boutsia20} in order to have a direct comparison of the performances of the PRF with other well-established techniques, e.g. the Canonical Correlation Analysis (CCA) used in the QUBRICS survey.

\subsection{The QUBRICS survey} \label{sec:QubricsSurvey}
The QUBRICS \textit{Main Sample} (hereafter \ms) contains objects with photometric measurements from three catalogues:
\begin{itemize}
	\item The $i$, $z$ magnitudes from the SkyMapper survey \citep[Data Release 1.1 ][]{ref:Skymapper};
	\item The G magnitude from the Gaia survey \citep[Data Release 2][]{ref:Gaia1, ref:Gaia2};
	\item The W1, W2, W3 magnitudes from the WISE survey \citep[][]{ref:WISE}.
\end{itemize}
In order to be included in the \ms{} an object must have a measured magnitude in {\it all} these six bands. Further additional constraints introduced in \citet{ref:Calderone19} are: ($i$) $14 < i_{\mathrm{psf}} < 18$, ($ii$) galactic latitude $|b_\mathrm{gal}|>25^\circ$, ($iii$) no photometric
flags in the $i$ and $z$ band\footnote{Photometric flags indicate issues during the image processing; pipeline specific flag are represented as power of two: 1, for instance, indicates that two sources are close enough to bias their respective photometry, 2 that distinct sources were initial blended, 4 the presence of saturated pixels in an object; the complete list is available in the SkyMapper DR1.1 documentation at this \href{http://skymapper.anu.edu.au/data-release/dr1/\#Coverage}{web-page}. Pipeline flags are combined with a bit-wise \texttt{OR} and the results are given in the published catalogue per source and photometric band, allowing end-users to exclude poorly processed objects \citep{ref:Skymapper}.}
($iv$) matching GAIA DR2 and WISE sources within 0.5" and ($v$) $\mathrm{SNR} > 3$ in the first three WISE bands; constraints have been designed to reduce contamination, leading to a total of 1014875 sources over approximately 12400 square degrees, mostly in the Southern Hemisphere. 
 
When available, additional data have been added: J, H and K$_s$ magnitudes from the 2MASS survey \citep[][]{ref:2mass}, \textit{u}, \textit{v}, \textit{g}, \textit{r} SkyMapper magnitudes, 
Gaia $G_{\texttt{RP}}$, $G_{\texttt{BP}}$ measurements and the W4 magnitude from WISE. GALEX \citep[][]{ref:GALEXBianchi17} data have been added by using the Mikulski Archive for Space Telescopes\footnote{the website is accessible at: \href{https://mast.stsci.edu/portal/Mashup/Clients/Mast/Portal.html}{MAST}.} (MAST). Sources have been cross-matched with a 5" matching radius; in the rare case of multiple matches the closest has always been retained. Only data produced as part of the All-Sky and Medium-Sky surveys (AIS and MIS respectively) have been selected.

\begin{figure}
    \centering
    \includegraphics[width=\linewidth]{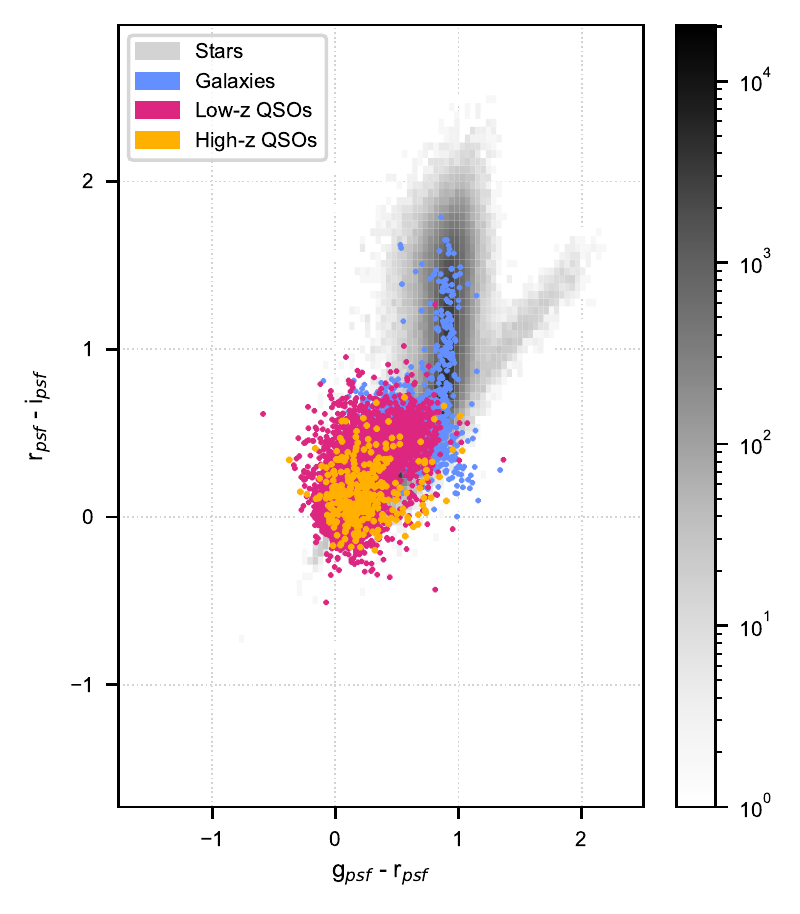}
    \caption{$r-i$ vs $g-r$ SkyMapper point spread function magnitudes for stars, galaxies, low and high redshift QSOs included in the Main Sample. While there are 4 visible clumps, there is not an efficient way to separate them. Stars were binned with an arbitrary bin-size to better visualize their distribution and the number counts for each bin is shown by the colormap on the right.}
    \label{fig:colColPlot}
\end{figure}

This additional photometric information is valuable to machine learning algorithms. As shown in fig. \ref{fig:colColPlot} and discussed in \citet[]{ref:carrasco15} it is not trivial to apply a simple colour-colour plot to separate QSOs, especially at high redshift, from contaminants (e.g. non-active galaxies or stars). Including additional photometric information such as infrared magnitudes from WISE helps in disentangling different populations, but it is not simple to devise appropriate color cuts in a multi-dimensional color space.

Parallax and proper motion information have been used to identify \textit{bona fide} stars: 83.1\% of sources in the \ms{} have been classified as such. The remaining entries have been matched with catalogues of known QSOs and extra galactic sources, in particular the SDSS DR14Q \citep[][]{ref:sdss}, the 13$^{th}$ edition of the Véron-Cetty catalogue \citep[][]{ref:Veron10} and the 2dFGRS \citep[][]{ref:2df}. The matching process identified 4666 confirmed QSOs and 3665 non-active galaxies. Matching against these catalogues, together with \textit{bona fide stars} identified through Gaia parallaxes and proper motion measurements, provided a source type classification for 84\% of the original \ms{}.
The remaining 16\% are unlabelled sources: as described in \citet{ref:Calderone19} and \citet{ref:Boutsia20} they are given an estimated classification (non-active galaxies, stars, low and high redshift QSOs) and redshift using the CCA. 

The CCA method produces a linear transformation matrix: when multiplied with an appropriate magnitude matrix a new label is obtained (hereafter \texttt{CCA}). The CCA procedure ensures that the \texttt{CCA} label is maximally correlated with the classification labels. The same transformation matrix, obtained on known sources, can be applied on unclassified objects: this allows to select the most favourable QSO candidates. The same procedure can be applied in order to obtain a redshift estimate and further exclude contaminants.

Extended objects were discarded to produce a sample of higher purity: this was accomplished following the same approach of \citet[]{ref:Calderone19}. A measure of the size of an object, \texttt{ext$_{iz}$} was obtained by comparing point spread function and petrosian magnitudes from the SkyMapper survey. The difference between the two was initially calculated for stars in the Main Sample, in order to derive a typical value, per magnitude interval, for point-like sources. The same quantity was then derived for unlabelled sources: those with \texttt{ext$_{iz}>3$} were excluded \textit{a priori} from the selection. In this way we consider it safe to use psf magnitudes as features, since our focus is on searching for point-like objects like quasars and most of the extended targets are excluded \textit{a priori}.

Finally, as part of the QUBRICS survey, various spectroscopic campaigns provided a secure identification for $\sim$500 targets, thus raising the number of known QSOs in the \ms{} to 5043; of these, 428 are at $z \geq 2.5$. Additional catalogue matching provided an identification for $\sim400$ non-active galaxies. The most recent number of sources in the \ms, used in this work, is shown in table \ref{tab:msSourceCount}.

\begin{table}
	\centering
	\begin{tabular}{cc}
		\toprule
		Source Type & Number \\
		\midrule
		All & 1014875 \\
		Unclassified sources & 162118 \\
		\textit{Bona-fide} Stars & 843690 \\
		Known non-active galaxies & 4024 \\
		Known QSOs (all $z$) & 5043 \\
		\midrule
		CCA QSO candidates (all) & 1412 \\ 
		CCA QSO candidates (not yet observed) & 818 \\
		\bottomrule
	\end{tabular}
	\caption{Number for all sources in the \ms, including the most recent QSO candidate sample.}
	\label{tab:msSourceCount}
\end{table}

\subsection{The Training Set} \label{sec:trainigSet}
Before applying the PRF to a selected dataset a few preliminary operations are needed:
\begin{itemize}
	\item in order to have a balanced number of stars, QSOs and galaxies for the training set we considered only a subset of all available stars.
	The latter have been chosen in order to evenly sample the available $i-z$ colour space: sources have been subdivided in bins (0.15 mag wide) based on their colour, and for each bin up to 600 stars where chosen. All objects in bins with less than 600 entries have been kept; bins with more than 600 entries have been randomly sampled: this produces a set of 5814 stars. The dataset built out of the 5814 stars, 5043 QSOs and 4024 galaxies will be referred to as \rms{} and will be the primary training sample for the PRF; 
	\item over-sampling the high redshift QSO sub-sample of the \rms{}. This is necessary to ensure an appropriate training test for the PRF when distinguishing high and low redshift sources. We used the \texttt{imbalanced-learn} python module \citep[][]{ref:Guillaume17UmbLearn}, in particular employing the \texttt{RandomOverSampler} method. This is the simplest oversampling method available: new samples are obtained by randomly drawing with replacement already available objects;
	\item distinguishing between nulls and non-detections: in particular an appropriate treatment of the flux upper limits can provide useful data to the algorithm, improving its predictive capabilities (\S\ref{sect:LimMag}). 
\end{itemize}

\subsection{Non-detections and missing data}
\label{sect:LimMag}

Missing data in photometric datasets are common and can be the result of two different occurrences: a measurement may be missing because a particular area of the sky has not been observed in a given pass-band or because the target is too faint to be detected in the pass-band. Despite apparently producing the same result in the final dataset - a missing value - the two cases should be treated differently in the implementation of the PRF algorithm, because the information content is different.

In the following, we will refer to missing data due to the first scenario as \n{} and to non-detections as \nd{}. If fed to the machine learning algorithm, both \n{} and \nd{} can provide additional information: in the following (\S\ref{sec:lowHighQSO}) it will be shown that supplying \nd{s} to the algorithm produces slightly higher completeness (from 77\% to 84\%) and lower contamination (from 28\% to 22\%).

In the PRF approach, \n{} are easily dealt with: given a node, for which the splitting condition (eq. \ref{eq:RFCondition}) is based on the $i^{th}$ feature, an object whose corresponding feature is a \n{} propagates to both left and right node with the same probability: 0.5 for the left, 0.5 for the right node. 

\nd{}, instead, should propagate like a measured feature, with an appropriate probability distribution.

In order to distinguish \nd{} and \n{} we have taken advantage of the additional information found in the published catalogues: SkyMapper and Gaia DR2, for instance, provide the number of visits per object per photometric band. If the number of visits is larger than zero, a missing value is considered a \nd, otherwise it is a \n. Often catalogues (e.g. WISE), already distinguish \nd{} and \n{}.

If a given catalog did not specify a limiting magnitude, we estimated
a reference value for \nd{} and an appropriate probability distribution from the properties of the catalog. In fact, for a given band,
objects counts are expected to increase as a function of the magnitude till incompleteness sets in (i.e. the probability of a non-detection becomes non-negligible and increases as a function of the magnitude until in practice it becomes one).

\begin{figure}
	\centering
	\includegraphics[width=\linewidth]{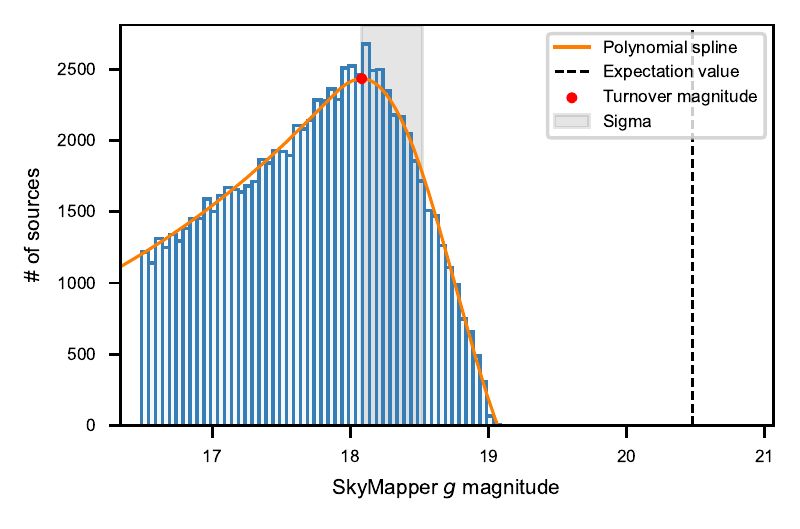}
	\caption{Magnitude distribution for the SkyMapper $g$ band (blue histogram). The yellow line represents the polynomial spline used to estimate the peak of the distribution (red dot at magnitude $\sim18$). The dashed, black line shows the reference value used for the $g$ band in the \rms, obtained as the expectation value of the appropriate PDF; the shaded area shows the $\sigma$ interval associated with the reference value.
	}
	\label{fig:limMag}
\end{figure}

We followed two different approaches:
\begin{enumerate}
\item{a non-detection is assumed to have a magnitude corresponding to a signal-to-noise ratio roughly equal to 1, to which a Gaussian PDF is associated with $\sigma=1.085$ (corresponding to SNR=1).
For example, in the case of the SkyMapper bands, assuming a background limited regime, we can determine the magnitude for which a typical SNR is achieved, e.g. $SNR=10$ (and, correspondingly, a $\sigma_m = 0.1085$), and from this magnitude, e.g. $m_{10}$, derive the reference value for \nd\ as:
\begin{equation}
    \texttt{ND} = m_{10} + 2.5 
\end{equation}
}
\item{the magnitude distribution at the limit of the detections has been used to estimate the probability distribution for
\nd\ (a low-pass distribution, eq. \ref{eq:lmDistribution}).
For each photometric band of interest, a large number ($\sim 10^5$ or more) of sources from the same survey has been collected from randomly selected regions in the Southern Hemisphere and a histogram has been built with a 0.05 magnitude bin. A polynomial spline has been used to interpolate the histogram and obtain a reliable estimate of the magnitude at the turnover of the counts (\texttt{TM}, the red dot in Fig.~\ref{fig:limMag}).
Once the \tm{} is determined, we introduced a low-pass distribution, expressed as:
\begin{equation}
    \label{eq:lmDistribution}
    \begin{cases}
        f(m) = N [1-\exp{(\frac{\tm - m}{\sigma}})]
    & \tm \leq m \leq \tm + \mathrm{k \sigma} \\
        f(m) = 0 & \mathrm{otherwise}
    \end{cases}
\end{equation}
where $\sigma$ is the 68\% percentile of the sources fainter than \tm, $k$ is the upper limit needed to have a finite distribution, and $N$ is the normalization coefficient needed to ensure that $ \int_{ \tm}^{\tm + \mathrm{k \sigma}} \mathrm{PDF}(m) ~ dm = 1 $. The parameter $k$ has been chosen to be 10, which produces values similar to those determined with  the approach {\it (i)}; 
the final non-detection value is obtained by calculating the expectation value of the distribution (\ref{eq:lmDistribution}) as $ \nd = \int_{\texttt{TM}}^{\texttt{TM} + k\sigma} m ~ \mathrm{PDF}(m) ~ dm$.
Different choices of $k$ do not affect significantly the results of the PRF classification, provided that its value is chosen large enough ($k > 5$). This is not surprising, since the PRF, in the training phase, has the capacity to adapt the probability thresholds described in Sect.~\ref{chap:prf} to our choice of $k$.
}
\end{enumerate}

Turnover values and standard deviations estimated for the \rms{} are listed in Table~\ref{tab:limMag}.

Both methods are rather rough approximations. They provide similar results
with slightly better results, in term of contamination of the test sample, for the approach (\textit{ii}), which has been adopted in the following.

\begin{table}
	\centering
	\begin{tabular}{cccc}\toprule
		Survey & Turnover & $\sigma$ \\
		band & magnitude (AB) & (mag) \\
		\midrule
		SkyMapper $u$ & 17.89 & 0.46 \\
		SkyMapper $v$ & 17.65 & 0.42 \\
		SkyMapper $g$ & 18.08 & 0.44 \\
		SkyMapper $r$ & 18.10 & 0.41 \\
		Gaia G$_{BP}$ & 20.81 & 0.46 \\
		Gaia G$_{RP}$ & 19.43 & 0.47 \\
		2MASS J & 16.55 & 0.29 \\
		2MASS H & 15.84 & 0.36 \\
		2MASS K$_s$ & 15.44 & 0.35 \\
		WISE W4 & 15.64 & 0.39 \\
		GALEX NUV (AIS) & 22.56 & 0.38 \\
		GALEX FUV (AIS) & 22.16 & 0.46 \\
		GALEX NUV (MIS) & 23.78 & 0.38 \\
		GALEX FUV (MIS) & 23.53 & 0.41 \\
		\bottomrule
	\end{tabular}
	\caption{Turnover, associated error $\sigma$ for each photometric band in the \rms{}.}
	\label{tab:limMag}
\end{table}

\section{Applying the PRF to the reduced main sample} \label{sec:results}

In this section we test the capabilities of the PRF in finding high-redshift (for our purposes $z>2.5$) QSOs, aiming at the production of a high-purity sample, in order to minimize the investment of telescope time for spectroscopic follow-up, but also of sufficient completeness for applications such as the calculation of luminosity functions. The algorithm will provide a classification based on the available magnitudes/colours, as the main source of information.

\subsection{General approach}
We first apply the PRF to a sample including all types of sources (i.e. stars, non-active galaxies and quasars of all redshifts). By construction, the \rms{} is built so that the fraction of each component is roughly one third of the total. Due to the typical surface densities of the various categories, the number of high redshift quasars with respect to the total is relatively small (roughly 8.5\% of all QSOs in the \rms). Special care will then be needed when dealing with predictions of the algorithm for this kind of objects. The number of sources for each class is listed in table \ref{tab:trainingNumbers}.

\begin{table}
	\centering
	\begin{tabular}{cc}\toprule
		Source Type & \# of sources available \\
		\midrule
		Quasar (all $z$) & 5043 \\
		Quasar ($z \geq 2.5$) & 428 \\
		Quasar ($z < 2.5$) & 4615 \\
		Non-Active Galaxies & 4024 \\
		Star & 5814 \\
		\bottomrule
	\end{tabular}
	\caption{Number counts for objects used in the \rms{}}
	\label{tab:trainingNumbers}
\end{table}

A training dataset for the PRF includes two components: a magnitude matrix (with associated errors) and a label vector, possibly with uncertainties in the classification. The latter are not mandatory, and in this work only feature (i.e. magnitude) errors have been used. Class labels are for the most part assigned by means of information derived from the literature or based on assumptions - for instance, \textit{bona fide} stars - and in both cases no uncertainties could be given. Two datasets have been used for the various tests: the \rms, and its sub-sample containing only QSOs. Classification labels (class-labels) for all tests are numerical: the association of a class with a number is completely arbitrary, and the results provided by the PRF do not depend on the choice of the label.

The algorithm is trained and validated using a $k$-fold cross validation; results are then checked against an independent test set. 
Validation+train and test dataset are randomly generated at each run using a defined random state; special care has been taken to ensure that objects in both the train and test datasets follow the same class distribution; the training+validation dataset has been chosen to be 80\% of the available sources; the remaining objects have been used as a test set. 

The PRF hyper-parameters, described in \citet[]{ref:reisPRF} and in the corresponding PRF GitHub repository, have been chosen on the basis of a 5-fold cross-validation test: a higher $k$ did non produce meaningful differences. Based on the results of the $k$-fold test we chose to use 200 trees for each test, \texttt{sqrt} for \texttt{max\_parameters} and 0.05 for \texttt{keep\_proba}; other parameters have been kept at their default values. Finally, each decision tree is built out of a bootstrapped sub-sample of the original training set (i.e. we set \texttt{bootstrap = True} during the PRF initialization). 
To avoid biases due to a small testing dataset we have also chosen to repeat the process 100 times: we have split the original dataset in training and testing using a defined random state - unique for each of the 100 iterations - and checked the consistency of the results.

The predictions produced by the algorithm have been evaluated on the basis of contamination (i.e. the complementary of the precision, the fraction of relevant instances among the retrieved instances) and completeness (i.e. the recall, the fraction of relevant instances that are retrieved), defined as: 

\begin{itemize}
	\item \textbf{Contamination}: the number of undesired, but selected, sources over the total number of selected sources. The definition can be restated as $\frac{\texttt{FP}}{\texttt{TP+FP}}$, where $\texttt{TP}$ is the number of true positives and $\texttt{FP}$ the number of false positives;
	\item \textbf{Completeness}: the ratio of the number of sources of interest identified by the algorithm over the total number of sources of interest (independently known). The definition can be restated as $\frac{\texttt{TP}}{\texttt{TP+FN}}$, where $\texttt{TP}$ is the number of true positives and $\texttt{FN}$ the number of false negatives. It should be noted that in this context the definition of completeness does not take into account ($i<18$) QSOs thay may be existing in the sky and are not present in the QUBRICS \ms\ because they were absent in some of the key databases, e.g. skymapper.
\end{itemize}

\subsection{The PRF as classifier} \label{sec:classifier}
\subsubsection{QSOs, stars and galaxies} \label{sec:qsostargal}
\begin{figure}
	\centering
	\includegraphics[width=\linewidth]{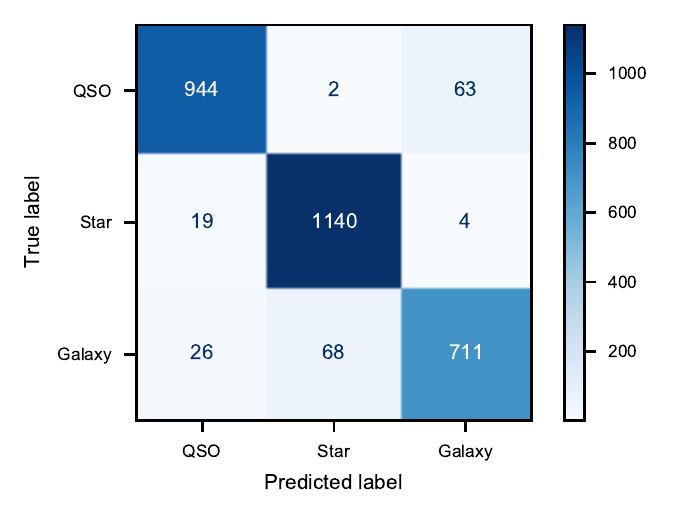}
	\includegraphics[width=\linewidth]{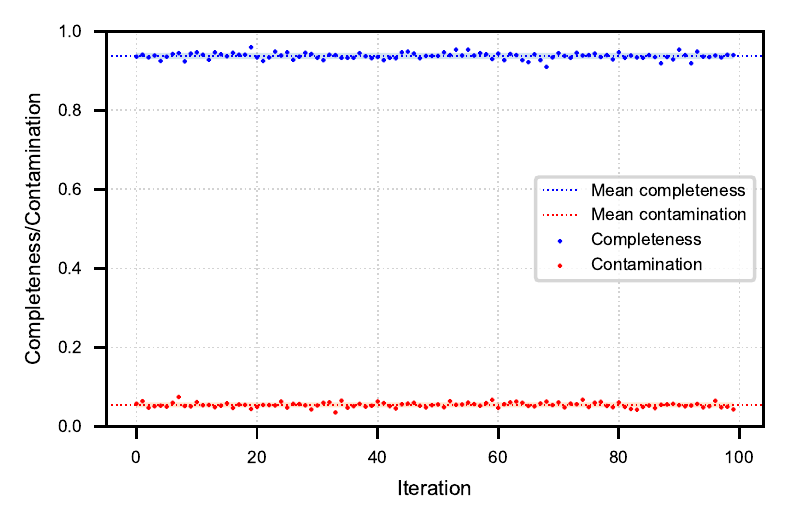}
	\caption{Top panel: Prediction of the algorithm for the test dataset when used to separate QSOs, stars and galaxies represented as a confusion matrix.
	Bottom Panel: completeness and contamination calculated over 100 iterations. The plot refers to QSOs only.} 
	\label{fig:QsoStarGalCM}
\end{figure}

The algorithm has been first used to distinguish QSOs, stars and non-active galaxies; the three classes have been labelled respectively as 1, 2 and 3. Training, validation and testing datasets have been all extracted from the whole \rms. A confusion matrix is used to visualize the prediction of the algorithm (fig. \ref{fig:QsoStarGalCM}, upper panel).

The test dataset is composed of 1009 QSOs, 1163 stars and 805 galaxies: as shown in fig. \ref{fig:QsoStarGalCM} roughly 93\% of QSOs are correctly classified by the algorithm. Moreover, most (62 out of 63) of misclassified QSOs are at $z<0.5$ and are identified as galaxies: this is not surprising, as the spectral energy distribution of low redshift quasars can be dominated by the host galaxy, and a spectroscopic observation is required to reveal the presence of the QSO. Both stars and galaxies contaminate the QSO sample, even if the latter are more commonly selected as QSOs. Considering the QSO as the target class the algorithm scores a recall of 93\% with a contamination of 5\%. We note however that our final aim is to identify high redshift ($z>2.5$) sources: out of the 89 in the test dataset, 88 ($\sim$99\%) are correctly classified by the algorithm as quasars.

This result is tied to a particular choice of train/testing datasets: different test sources might provide slightly different results; we thus evaluated the performance of the algorithm 100 times, using a different test dataset at each iteration, averaging the results at the end. Results are shown in fig. \ref{fig:QsoStarGalCM} lower panel, where both completeness and contamination are calculated with respect to the QSO class: the average completeness (contamination) is 93.7\% (5.5\%) with a scatter of $\sigma = 0.8\%$ ($\sigma = 0.7$\%); the percentage of high redshift QSOs identified by the algorithm is, in each run, $\sim98.5\%$ with a large scatter, $\sim 1.5\%$.

\subsubsection{Low and high redshift QSOs} \label{sec:lowHighQSO}
As the final goal of this work is to identify QSOs with $z>2.5$, having a reliable classification as a QSO is not sufficient: we still need to exclude low-redshift QSOs that in \citet{ref:Calderone19} and \citet{ref:Boutsia20} have been found to be the major contaminant.
To this end we have applied a second time the PRF to the objects classified as QSOs in the previous step, trying to discriminate whether their redshift is higher or lower than a threshold, initially chosen to be $z = 2.5$ in accordance with the QUBRICS definition of high-$z$ QSOs.
This choice results in an unbalanced training dataset (as shown in tab. \ref{tab:trainingNumbers}), as the number of objects above $z = 2.5$ is just the 8.5\% of the total. Moreover, more than 2/3 of sources available for training at $z < 2.5$ are at $0 < z < 1.5$: in both cases the redshift distribution of sources in the training sample is not suitable for our purposes and negatively impact on the performances of the algorithm. 
In order to mitigate the issue we have chosen to apply a simple oversampling method: in our approach, new samples are generated by sampling with repetition the available data. We experimented with different oversampling strategies, in order to find the best compromise in term of completeness vs contamination. The best results were obtained by applying the over-sampling algorithm twice: once for objects with redshift between 0 and 3, in order to produce a ratio of sources with $2 \leq z < 3$ to those with $0 \leq z < 2$ equal to 0.5; the second for sources at $z \geq 2.5$, in order to match the number of low redshift objects.
This choice has been adopted to preserve all the available information in the dataset: under-sampling the majority class would remove precious information which could instead be used by the algorithm. Moreover, we have not used a more advanced over-sampler \citep[e.g. SMOTE ][]{ref:smote} due to the missing values in our dataset: SMOTE creates synthetic instances by searching for nearest neighbours and averaging over their corresponding feature values. 

\noindent These results also suggest that the dataset currently available is small with respect to the complexity of the problem: a better and more ample training set would greatly benefit the performances of the PRF.

\begin{figure}
	\centering
	\includegraphics[width=0.85\linewidth]{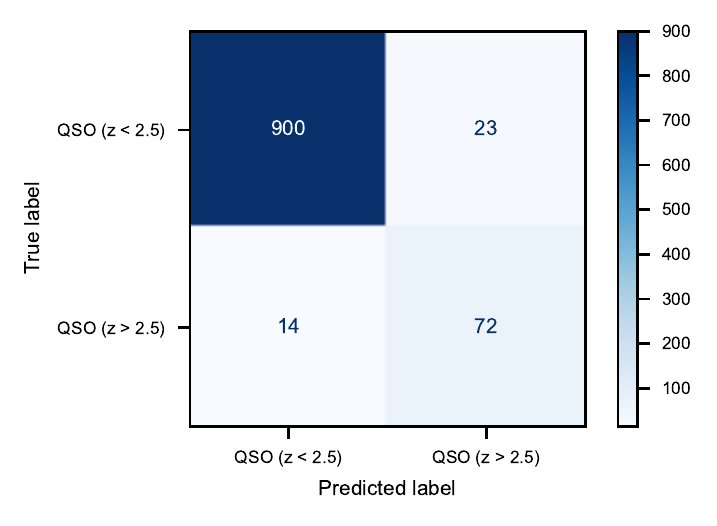}
	\includegraphics[width=\linewidth]{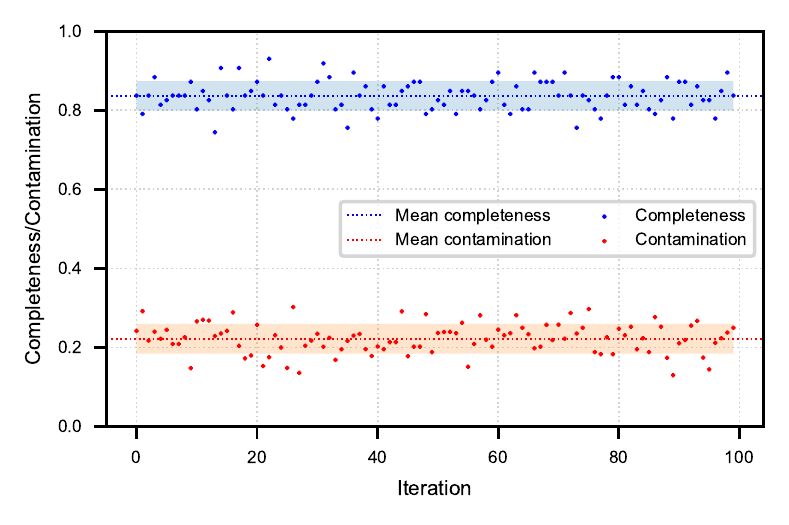}
	\caption{Results of the classification process high vs. low redshift QSOs in the case of a redshift threshold of $z = 2.5$. Top panel: confusion matrix for this test. Bottom panel: completeness and contamination calculated over 100 runs with different test and train dataset. Dashed lines mark the average values, dots results for a particular run while the shaded regions denote the 1-$\sigma$ interval.}
	\label{fig:lowHighQsoCM}
\end{figure}

The available data have been subdivided once again in training+validation and testing, with the same ratio used in the previous test (80\% - 20\%). Results for the classification process on a test dataset are shown in fig. \ref{fig:lowHighQsoCM}.

We repeated the same procedure described in the previous section, in order to avoid biases due to the particular training/testing sample. In this case we achieve an average completeness (contamination) of $\sim$84\% ($\sim$22\%), with higher scatter with respect to what we observed before: 3\% for both completeness and contamination. These values can be compared to those obtained for the QUBRICS survey for a similar threshold \citep[fig. 5 in][green line]{ref:Calderone19}: the \texttt{CCA} produces a dataset with a slightly higher completeness, but higher contamination ($\sim$37\%). 

\begin{figure}
	\centering
	\includegraphics[width=\linewidth]{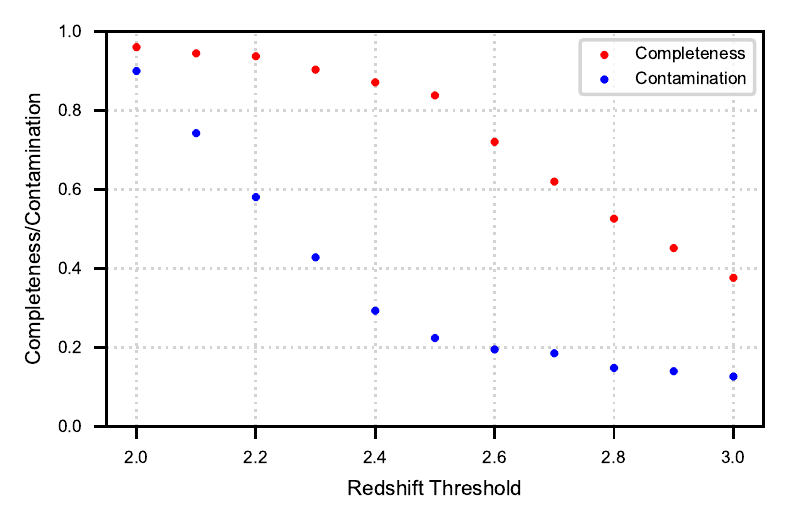}
	\caption{Completeness (red dots) and contamination (blue dots) as a function of the delimiting redshift for the low and high QSO classes. Each completeness-contamination pair in the plot is the average of 100 iterations on different test datasets, and is calculated with respect to the $z = 2.5$ threshold.}
	\label{fig:completenesscontamination}
\end{figure}

During the training process it is possible to set a lower redshift threshold in order to produce higher completeness - at the expense of higher contamination - with respect to the original, $z = 2.5$ threshold. Conversely, higher redshift thresholds will produce lower contamination and, at the same time, lower completeness. In order to test the effect on the selection process different thresholds have been selected sampling the redshift interval $z = 2-3$ with steps of 0.1 in redshift units. The results are shown in fig. \ref{fig:completenesscontamination}. 

As expected, both completeness and contamination rise as the threshold value becomes lower. In order to obtain the same completeness expected for QUBRICS, one should take a redshift limit of $z\simeq2.3$; the contamination is similar to that achieved with the CCA (roughly 40\%). The completeness in the test sample rapidly decreases (40\% from $z = 2.5$ to $z = 3.0$) as redshift thresholds get higher, whereas the contamination decreases at a slower pace (roughly 10\% in the same redshift interval): we thus chose to use the $z = 2.5$ threshold in the application on the unclassified sample.

\subsubsection{Analysis of the contaminants}
The contaminants affecting the final sample of high-redshift QSOs will be of two types:
lower-redshift ($z < 2.5$) QSOs wrongly classified at high redshift and galaxies/stars misclassified as QSOs in the first step that
survive the subsequent classification as high-redshift QSOs.

The performance of the algorithm for the first type of contaminants has been treated in the previous subsection. The probability of a
QSO with $z<2.5$ to be classified as a high-z QSO ($PQ$ in the following), as shown in Fig.~\ref{fig:lowHighQsoCM}, is on average $2\%$.

To quantify the global performance concerning non-QSO contaminants it is necessary to combine the two previous tests. 
To this end, the algorithm has been initially trained on part (80\%) of the \rms{} and applied on a test dataset (the remaining 20\%); both train and test datasets contain stars, galaxies and QSOs, and this produces a QSO candidate sample at all redshifts, which includes misclassified sources, i.e. stars and galaxies predicted to be QSOs. These are then re-classified as low or high redshift sources, allowing to verify how many non-QSO contaminants are picked up at the end by the algorithm. 
The process has been repeated 100 times, each with different train-test datasets. 

On average 4\% ($\sim35$)
of the galaxies and 2\% ($\sim 20$) of the stars are classified as generic QSOs. 
Of these, 0.1\% of the galaxies ($PG$ in the following) and 0.1\% (1) of the stars on average are classified as high redshift sources. 

In order to avoid issues tied to the small number of objects in the test set, we repeated the process using  all \textit{bona fide} stars in the \ms{}, excluding those part of the \rms{} and used during the training process. Out of the 837876 initial \textit{bona fide} stars, on average $\sim2400$ (0.3\%) are picked up by the algorithm as generic QSOs at all redshifts. 
Of these only $\sim400$ (0.05\%, $PS$ in the following) are selected as QSO candidates at $z\geq2.5$. 

\section{Characterization and Comparison with the CCA selection} \label{sec:characterization}
In order to obtain a list of high-redshift QSO candidates, the PRF has first been applied on the unclassified sources in the \ms,
trained as described in the section \ref{sec:qsostargal}.
As described in section \ref{sec:QubricsSurvey}, extended objects have been discarded \textit{a priori},
leaving a total of 58782 objects (Unclassified Dataset, \texttt{UD}).

The PRF, applied to the \texttt{UD}, has
produced a list of 22113 QSO candidates (at all redshifts), 18573 stars and 18096 potential galaxies.

In order to select high-redshift ($z\geq2.5$) QSOs
the PRF has been trained on the QSO sub-sample of the \rms{}, as described in section \ref{sec:lowHighQSO}, and then has been applied to the previously selected generic QSO candidates. The threshold defining the high- vs. low-redshift class has been chosen at $z = 2.5$. This second selection has identified a final sample of 626 high-$z$ QSO.
\begin{figure}
    \centering
    \includegraphics[width=\linewidth]{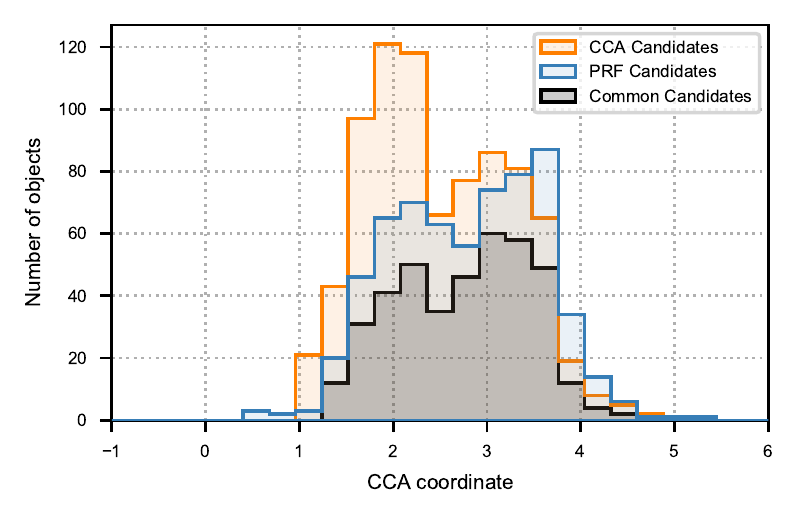}
    \includegraphics[width=\linewidth]{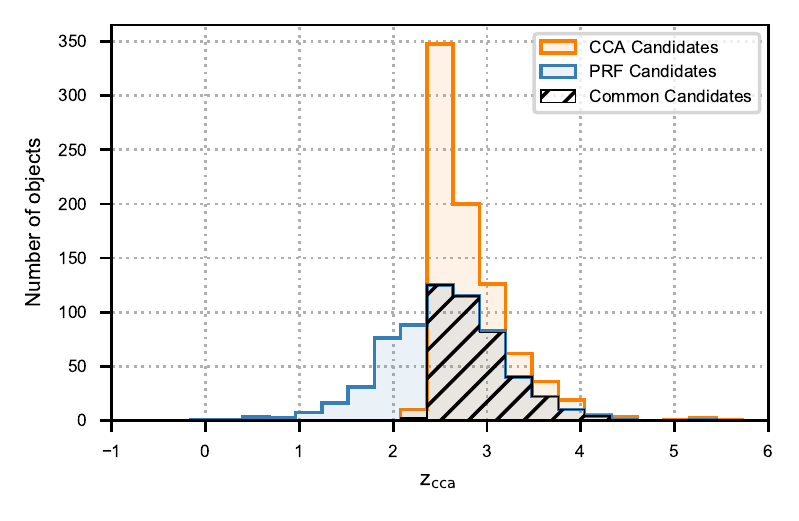}
    \caption{\texttt{CCA} coordinate (top panel) and CCA redshift estimate ($z_\texttt{CCA}$ (bottom panel) for sources selected by the PRF (blue), CCA (orange) or both algorithms (green). Five sources at $z_\texttt{CCA} > 10$ have been excluded from the plots, and are most likely contaminants.
    The \texttt{CCA} coordinate corresponds to an object-type classification, while the $z_\texttt{CCA}$ is an estimate for the redshift of the object. As described in \citet[]{ref:Calderone19}, non active galaxies are spread around \texttt{CCA} $\sim-1$, stars around \texttt{CCA} $\sim$ 0, low-$z$ QSOs around \texttt{CCA} $\sim2$ and high redshift QSOs have $z_\texttt{CCA}\gtrsim3$.
    \label{fig:CCAPRFCompare}
    }
\end{figure}

The completeness of this list, for $z\ge 2.5$ quasars is expected to be, according to section \ref{sec:classifier}, $83\%$.

In order to estimate the success rate of a spectroscopic follow-up we can assume that the partition of the unclassified sources in the \ms\ in stars, galaxies and generic quasars is described by the results of the PRF classifier, i.e. 31\% of stars, 31\% of galaxies, and  38\%  of generic QSOs.
The number of expected contaminants, due to the misclassification of stars and galaxies,
turns out to be low:
$ 18573 \times PS = 18 $ and $18096 \times PG = 9$,
respectively.

To estimate the more significant contamination of misclassified low-z QSOs and, conversely,
the number of high-z QSOs not selected,
we have convoluted the expected redshift distribution of $i \le 18$ QSOs, derived from 
\cite{Shen2020}, with the probability, as a function of the redshift, for a $z<2.5$ QSO to be classified at high-z and for a $z\ge 2.5$ QSO to be classified at low-z (computed as in Section \ref{sec:classifier}).

As a result, in the list of 626 high-z QSO candidates we expect to have
about 66\% (411) true  $z\ge 2.5$ QSOs, 30\% (188) $z<2.5$ QSOs, 4\% (27) galaxies or stars.
16\% of the $z\ge 2.5$ QSOs, mainly around $z=2.5-2.9$, are expected to have been missed in the selection. The lower-z QSO misclassified as high redshift sources with $z>2.5$, are actually expected to have a redshift $<z> \sim 2.1$ and all at $z>1.5$.

\subsection{\textbf{Comparison with the CCA predictions}}

The sample of candidates obtained using the PRF has been compared with the equivalent list produced using the CCA method \citep{ref:Calderone19}: 401 sources turned out to be in common, while 417 and 225 objects are exclusively selected by the CCA and PRF respectively. Figure \ref{fig:CCAPRFCompare} shows the comparisons between the \texttt{CCA} coordinate (top panel) and the $z_\texttt{CCA}$ estimate for PRF (CCA) and common candidates.
Most (> 95\%) of the PRF selected sources are at $\texttt{CCA} > 1$, corroborating the choice made in \citet{ref:Calderone19} to exclude from the candidate list unclassified sources with $\texttt{CCA}<1$. 
On the other hand, a significant (190 out of 227)
part of PRF selected objects is at $z_\texttt{CCA}<2.26$: 61\% of these are however within the 1-$\sigma$ scatter found for the \texttt{CCA} selection estimate.
Taking the corresponding estimates for the completeness and success rate of the CCA method from \cite{ref:Boutsia20}, and with the rough assumption that the CCA and PRF selection are statistically independent, we would expect the intersection of the two selections to have a 89\% success rate and a 75\% completeness.

\subsection{Colour comparison}
The PRF uses magnitudes as features to provide a label for unclassified objects. It is interesting to compare the magnitude or colour distribution of the newly classified objects to that of the training: we expect the colour distribution of the two to be similar. The $i - z$ colour was chosen in this case, as all sources in the \ms{} are required to have a reliable magnitude in these two bands.

\begin{figure}
    \centering
    \includegraphics{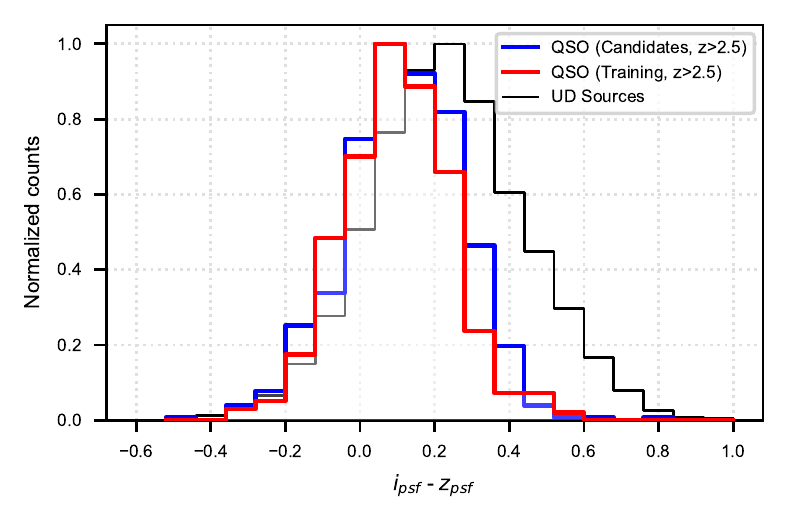}
    \caption{The $i_{psf} - z_{psf}$ colour for QSOs at $z>2.5$. The red histogram shows to the colour distribution for objects in the training sample, the blue one for candidates and the grey one for all unclassified sources (UD dataset). Each histogram has been normalized by its maximum in order to show all of them in the same plot.}
    \label{fig:colComparison}
\end{figure}

As shown in fig. \ref{fig:colComparison} there is good agreement between the distribution of colours for the training and candidate sample, especially for high redshift QSOs, the main interest for this work. This holds true for non-active galaxies as well, while the relation is less tight with stars: the training set was built to fully sample the available $i-z$ colour space, while the \texttt{UD} occupies a slightly smaller range.

The effect of \texttt{Nulls} was verified as well, considering sources with at least one \texttt{Null} in a photometric band: these are rather scarce, and the results consequently noisy. Due to their small number and random nature (e.g., a switched-off detector or a corrupted image), \texttt{Nulls} are not expected to produce significant deviations in the colour distributions. Considering the SkyMapper $u_{psf}$ as an example (one of the bands where \texttt{Nulls} are more numerous), there are 26 of 5043 QSOs in the training sample and 215 of 22113 QSO candidates with a \texttt{Null} value. The sample mean $i-z$ for QSOs in the training sample (QSO candidates) is 0.08 (0.13) with a standard deviation of 0.17 (0.18): hence, there is not a significant difference in the two distributions. 

\section{Spectroscopic Validation} 
\label{sec:spectroscopy}
Since the creation of the PRF candidate list,
observations of 41 sources
have been carried out at Las Campanas Observatory and at Telescopio Nazionale Galileo (TNG, La Palma) using LDSS-3 (Clay Telescope), IMACS (Baade Telescope) and Dolores. 

Observations at LDSS-3 have been obtained with the VPH-all grism, 1"-central slit and no blocking filter, covering a wavelength range between 4000 and 10000\AA{} and a low ($\sim$800) resolution. Observations at IMACS used the \#300 grism with a 17.5deg blaze angle with a dispersione of 1.34 \AA/pixel and the same wavelength range of LDSS-3. Exposures at TNG have been taken during the AOT41 and AOT42 periods under two proposals (PI: G. Calderone and A. Grazian); the LR-B grism with a 1" slit aperture have been used. 

Out of the 41 PRF selected and observed sources,
29 turned out to be genuine high-$z$ ($z>2.5$) QSOs and 12 QSOs with $1.88 < z < 2.5$; no stars nor galaxies have been selected by the algorithm. The results of the spectroscopic observations are summarized in table \ref{tab:newSpecObs}. 
In these preliminary observations we achieved a success rate of $\sim$70\%, 
that becomes 80\% if we consider the candidates in common with the CCA selection.
The $z<2.5$ contaminant QSOs turn out to be 12, with an average redshift $<z> = 2.2$ 
and a minimum redshift of z=1.888 (for a BAL QSO, ID=1044577), in good agreement with the predictions of Sect.~\ref{sec:characterization},
based on the characterization of our selection method.
As observed in \citet{ref:Boutsia20} and detailed in \citet{ref:Cupani2021} lower-z, extremely strong BAL QSOs are picked up because their huge absorption troughs tends to mimic the colors of higher-redshift QSOs.

It should be noted that the number of observed targets is still low ($\sim5\%$ of the whole candidate list), and in these exploratory runs targets have not been chosen in a systematic way and might not be entirely representative of the final performance of the PRF method.
In any case, the results appear encouraging and further observations worth pursuing,
possibly in parallel with other selection techniques, in order to better evaluate the capabilities of the PRF method and enlarge and make more complete the sample of bright high-redshift quasars in the Southern Hemisphere.

\begin{table*}
    \centering
    \begin{tabular}{cccccccccc}
    \toprule
     & QUBRICS & R.A. & DEC. & $m_i$ & $z_{\rm spec}$ &  Class & Obs. Date & Instrument & CCA \\ 
     & ID & \multispan2 {(J2000)} & (AB mag) & & & & & selected \\
    \midrule
    1  & 814160	    & 13:01:18.31	& -08:10:14.81	& 17.68	& 3.281	& QSO	& 2021-01-29	& LDSS3	    & Y \\
    2  & 824362	    & 12:33:22.24	& -11:53:39.53	& 17.90	& 3.183	& QSO	& 2021-01-29	& LDSS3	    & Y \\
    3  & 831008	    & 20:43:56.68	& -00:39:08.48	& 17.86	& 2.894	& QSO	& 2020-10-23	& DOLORES   & Y \\
    4  & 831970	    & 20:44:59.66	& -02:54:38.28	& 17.79	& 2.851	& QSO	& 2020-10-23	& DOLORES   & Y \\
    5  & 842834	    & 12:11:20.09	& -33:14:27.46	& 17.73	& 3.826	& QSO	& 2021-01-29	& LDSS3	    & Y \\
    6  & 859489	    & 21:58:00.41	& -07:18:05.50	& 17.80	& 2.538	& QSO	& 2020-10-23	& DOLORES	& Y \\
    7  & 859798	    & 21:33:26.01	& -03:23:32.70	& 17.78	& 2.518	& QSO	& 2020-10-23	& DOLORES	& N \\
    8  & 861290	    & 21:31:58.81	& -04:54:39.47	& 17.97	& 3.381	& QSO	& 2020-10-23	& DOLORES	& Y \\
    9  & 861881	    & 21:54:15.85	& -04:45:21.89	& 17.65	& 2.366	& QSO	& 2020-09-13	& DOLORES	& Y \\
    10 & 864254	    & 23:34:54.76	& -69:30:42.84	& 17.86	& 3.894	& QSO	& 2021-01-02	& IMACS	    & Y \\
    11 & 866799*	& 02:10:51.46	& -84:54:37.57	& 17.17	& 3.685	& QSO	& 2020-11-27	& IMACS	    & Y \\
    12 & 893444	    & 02:53:56.09	& -20:26:39.68	& 17.86	& 3.022	& QSO	& 2021-01-28	& LDSS3	    & Y \\
    13 & 908786	    & 23:22:27.67	& -04:48:45.16	& 17.78	& 2.589	& QSO	& 2020-10-23	& DOLORES	& Y \\
    14 & 992797*	& 01:07:10.57	& -62:36:48.35	& 17.28	& 2.833	& QSO	& 2020-10-07	& LDSS3	    & Y \\
    15 & 995059	    & 23:32:46.67	& -08:23:44.03	& 17.31	& 2.894	& QSO	& 2020-10-23	& DOLORES	& Y \\
    16 & 1005746	& 04:14:29.34	& -05:56:14.38	& 17.33	& 2.417	& QSO	& 2020-09-13	& DOLORES	& N \\
    17 & 1007009*   & 15:07:26.88   & -16:25:42.30  & 17.86 & 3.015 & QSO   & 2021-02-27    & IMACS     & Y \\
    18 & 1013347	& 04:42:30.12	& -26:32:19.05	& 17.55	& 2.914	& QSO	& 2020-11-27	& IMACS	    & Y \\
    19 & 1018840	& 03:27:24.51	& -52:38:58.20	& 17.79	& 3.771	& QSO	& 2020-12-31	& IMACS	    & Y \\
    20 & 1026400	& 06:36:44.21	& -63:40:33.04	& 17.69	& 2.410	& QSO	& 2020-11-24	& LDSS3	    & Y \\
    21 & 1030917	& 00:36:25.37	& -32:23:36.55	& 17.80	& 3.512	& QSO	& 2020-11-26	& LDSS3	    & Y \\
    22 & 1031280	& 22:33:47.55	& -04:02:04.60	& 17.89	& 2.092	& QSO	& 2020-10-23	& DOLORES	& N \\
    23 & 1031462	& 22:36:06.12	& -16:10:34.21	& 17.80	& 2.109	& QSO	& 2020-10-23	& DOLORES	& Y \\
    24 & 1031929	& 02:10:25.34	& -38:17:17.96	& 17.42	& 3.308	& QSO	& 2020-11-26	& LDSS3	    & Y \\
    25 & 1032609	& 00:33:11.87	& -40:51:49.35	& 17.82	& 1.963	& QSO	& 2020-11-26	& LDSS3	    & Y \\
    26 & 1033197	& 23:41:33.99	& -20:24:08.81	& 17.56	& 2.603	& QSO	& 2020-09-13	& DOLORES	& Y \\
    27 & 1034040	& 00:58:23.18	& -09:04:34.98	& 17.82	& 2.775	& QSO	& 2020-10-23	& DOLORES	& Y \\
    28 & 1034851	& 01:18:08.11	& -23:07:56.31	& 17.36	& 3.096	& QSO	& 2020-10-08	& LDSS3	    & Y \\
    29 & 1035092	& 00:29:55.80	& -22:26:28.53	& 17.73	& 2.782	& QSO	& 2020-10-23	& DOLORES	& Y \\
    30 & 1039886	& 22:40:56.35	& -08:03:58.41	& 17.71	& 2.463	& QSO	& 2020-09-13	& DOLORES	& N \\
    31 & 1040503	& 01:24:36.61	& -31:26:23.61	& 17.20	& 1.902	& QSO	& 2020-10-07	& LDSS3	    & Y \\
    32 & 1041074	& 00:55:53.34	& -23:07:43.84	& 17.90	& 2.218	& QSO	& 2020-10-08	& LDSS3	    & N \\
    33 & 1041119	& 23:09:35.13	& -15:13:14.27	& 17.95	& 2.339	& QSO	& 2020-10-23	& DOLORES	& N \\
    34 & 1044054	& 00:31:50.88	& -18:20:21.84	& 17.75	& 3.519	& QSO	& 2020-10-23	& DOLORES	& Y \\
    35 & 1044577*	& 00:21:11.30	& -17:29:01.04	& 17.68	& 1.888	& QSO	& 2020-09-13	& DOLORES	& Y \\
    36 & 1059422	& 01:54:48.05	& -10:49:40.61	& 17.43	& 2.556	& QSO	& 2020-10-08	& LDSS3	    & Y \\
    37 & 1080395	& 03:12:52.40	& -31:38:33.21	& 17.83	& 3.879	& QSO	& 2020-11-27	& IMACS	    & Y \\
    38 & 1086629	& 04:39:25.68	& -43:49:17.87	& 17.68	& 3.516	& QSO	& 2021-01-31	& IMACS	    & Y \\
    39 & 1094391	& 04:55:55.50	& -64:58:35.35	& 17.48	& 2.444	& QSO	& 2021-01-31	& IMACS	    & Y \\
    40 & 1101726    & 14:59:01.01   & -02:51:05.79  & 17.75 & 3.354 & QSO   & 2021-02-27    & IMACS     & Y \\
    41 & 1122453	& 02:35:57.55	& -34:48:56.45	& 17.79	& 3.737	& QSO	& 2020-11-24	& LDSS3	    & Y \\
    \bottomrule
    \end{tabular}
    \caption{Observed, PRF selected sources with a robust spectroscopic redshift estimate. Sources with a * show strong Broad Absorption Lines (BAL QSOs).}
    \label{tab:newSpecObs}
\end{table*}
    
\section{Conclusions} \label{sec:conclusions}
Searching for QSOs is a challenging task, even more so if relatively high-redshift sources are the goal. In this paper we presented a selection method based on a machine learning algorithm, the Probabilistic Random Forest, and used it to select relatively bright ($i<18$) high-redshift ($z>2.5$) QSOs. The PRF has been applied to the same initial dataset used for the QUBRICS survey \citep{ref:Calderone19}, including photometric estimates from the SkyMapper DR1, Gaia DR2, WISE, 2MASS and GALEX surveys.
We have first used the PRF algorithm to select QSOs (at all redshifts), in order to remove stars and non-active galaxies; we then re-classified the  QSO candidates, in low- and high-z QSO candidates. Our tests show that,
when applied to the QUBRICS sample, the PRF selection has a completeness of $\sim 83\%$ in selecting high-redshift sources, with a relatively low contamination of $\sim$22\%. Similarly to what observed in \citet[]{ref:Calderone19}, the main responsible for contamination turn out to be low-$z$ QSOs (93\%); stars and non-active galaxies are of secondary importance ($\sim 3$\% and $\sim 4\%$ respectively).

When applied to the unclassified dataset of QUBRICS, which contains 58782 sources, the algorithm produces a list of 626 high redshift QSO candidates: of these, 401 are in common with the equivalent CCA sample of \citet{ref:Calderone19}, while the remaining 225 are exclusively selected by the PRF. 

With preliminary observations of 41 PRF candidates we have been able to 
confirm 29 new high-$z$ sources, with a success rate close to our expectations. 
Further spectroscopic identifications are needed to better assess the capabilities of the PRF method. 

The relatively small number of high-redshift QSOs available for the training (< 10\% of the total) likely hampers the PRF performances, and we have had to resort to oversampling in order to obtain
reasonably uniform training sets, as described in Sect. \ref{sec:classifier}. Nonetheless, the PRF has proven to be a powerful and flexible technique to select high-redshift quasars, competitive with respect to
other techniques such as the CCA.

We are refining the selection methods and continuing the spectroscopic campaigns, 
in order to further improve the completeness and success rate of the QUBRICS survey
and to extend, with the growth of the training sets, the predictive capabilities to more specific QSO categories (e.g. \citet{Boutsia_2021}, and Cupani et al., in preparation).

\section*{Acknowledgements}
We thank Societ\`a Astronomica Italiana (SAIt), Ennio Poretti, Gloria Andreuzzi, Marco Pedani, Vittoria Altomonte and Andrea Cama for the observation support at TNG. Part of the observations discussed in this work are based on observations made with the Italian Telescopio Nazionale Galileo (TNG) operated on the island of La Palma by the Fundacion Galileo Galilei of the INAF (Istituto Nazionale di Astrofisica) at the Spanish Observatorio del Roque de los Muchachos of the Instituto de Astrofisica de Canarias.

The national facility capability for SkyMapper has been funded through ARC LIEF grant LE130100104 from the Australian Research Council, awarded to the University of Sydney, the Australian National University, Swinburne University of Technology, the University of Queensland, the University of Western Australia, the University of Melbourne, Curtin University of Technology, Monash University and the Australian Astronomical Observatory. SkyMapper is owned and operated by the Australian National University’s Research School of Astronomy and Astrophysics. The survey data have been processed and provided by the SkyMapper Team at ANU. The SkyMapper node of the All-Sky Virtual Observatory (ASVO) is hosted at the National Computational Infrastructure (NCI). Development and support the SkyMapper node of the ASVO has been funded in part by Astronomy Australia Limited (AAL) and the Australian Government through the Commonwealth’s Education Investment Fund (EIF) and National Collaborative Research Infrastructure Strategy (NCRIS), particularly the National eResearch Collaboration Toolsand Resources (NeCTAR) and the Australian National Data Service Projects (ANDS).

This work has made use of data from the European Space Agency (ESA) mission Gaia (https://www.cosmos.esa.int/gaia), processed by the Gaia Data Processing and Analysis Consortium (DPAC, https://www.cosmos.esa.int/web/gaia/dpac/consortium). Funding for the DPAC has been provided by national institutions, in particular the institutions participating in the Gaia Multilateral Agreement.

This publication makes use of data products from the Two Micron All Sky Survey, which is a joint project of the University of Massachusetts and the Infrared Processing and Analysis Center/California Institute of Technology, funded by the National Aeronautics and Space Administration and the National Science Foundation. 

This publication makes use of data products from the Wide-field Infrared Survey Explorer, which is a joint project of the University of California, Los Angeles, and the Jet Propulsion Laboratory/California Institute of Technology, funded by the National Aeronautics and Space Administration. 

This paper includes data gathered with the 6.5 meter Magellan Telescopes located at Las Campanas Observatory, Chile.

This research is based on observations made with the Galaxy Evolution Explorer, obtained from the MAST data archive at the Space Telescope Science Institute, which is operated by the Association of Universities for Research in Astronomy, Inc., under NASA contract NAS526555.

\section*{Data Availability}
The data underlying this article will be shared on reasonable request to the corresponding author.



\bibliographystyle{mnras}
\bibliography{bib}





\bsp	
\label{lastpage}
\end{document}